\documentclass[lettersize,journal]{IEEEtran}
\usepackage{amsmath,amssymb,mathtools}
\usepackage{bm}
\usepackage{algorithmic}
\usepackage{algorithm}
\usepackage{array}
\usepackage[caption=false,font=footnotesize]{subfig}
\usepackage{textcomp}
\usepackage{url}
\usepackage{verbatim}
\usepackage{graphicx}
\usepackage{cite}
\usepackage{microtype}
\usepackage{siunitx}

\newcommand{\R}{\mathbb{R}}
\newcommand{\N}{\mathbb{N}}
\newcommand{\E}{\mathbb{E}}
\newcommand{\Rge}{\R_{\ge 0}}
\newcommand{\relu}{\operatorname{ReLU}}

\newcommand{\abs}[1]{\left|#1\right|}

\newcommand{\set}[1]{\left\{#1\right\}}

\begin{document}

\title{A Power Electronic Converter Control Framework Based on Graph Neural Networks --- An Early Proof-of-Concept}

\author{Darius Jakobeit and Oliver Wallscheid,~\IEEEmembership{Senior Member,~IEEE}% <-this % stops a space
  \thanks{D. Jakobeit is with the Department of Power Electronics
    and Electrical Drives at Paderborn University, Germany. O. Wallscheid is
    with the Department of Interconnected Automation Systems at University of Siegen, Germany. E-mail: jakobeit@lea.uni-paderborn.de,
    oliver.wallscheid@uni-siegen.de}%
  %\thanks{This paper was produced by the IEEE Publication Technology Group. They are in Piscataway, NJ.}% <-this % stops a space
  %\thanks{Manuscript received April 19, 2021; revised August 16, 2021.}
}



\maketitle

\begin{abstract}
  Power electronic converter control is typically tuned per topology, limiting transfer across heterogeneous designs. This letter proposes a topology-agnostic meta-control framework that encodes converter netlists as typed bipartite graphs and uses a task-conditioned graph neural network backbone with distributed control heads. The policy is trained end-to-end via differentiable predictive control to amortize constrained optimal control over a distribution of converter parameters and reference-tracking tasks. In simulation on randomly sampled buck converters, the learned controller achieves near-optimal tracking performance relative to an online optimal-control baseline, motivating future extension to broader topologies, objectives, and real-time deployment.
\end{abstract}


%==========================
\section{Introduction}
%==========================

Power electronic converters (PECs) cover a vast range of circuit topologies, load characteristics and application tasks. This problem landscape must be matched by appropriate control strategies. Hence, controller-synthesis automation methods have gained significant attention in recent years to reduce engineering effort. Notable examples are self-commissioning data-driven predictive control such as DeePC \cite{huang2019deepc_converter}, supervised learning to approximate predictive control laws \cite{novak2021imitation}, and reinforcement learning (RL) including transfer/meta-learning for wide parameter ranges \cite{jakobeit2023metaRL_pmsm}. These approaches significantly reduce tuning effort, but they usually assume a fixed PEC topology and at best a limited range of varying parameters.

This yields an open research gap: topology-agnostic meta optimal control that transfers across heterogeneous converters and tasks, i.e., going beyond parameter adaptation. The central idea of this letter is to represent any converter by a graph and to use a graph neural network (GNN) as a permutation-consistent encoder (backbone) that produces a shared latent representation for downstream control \cite{battaglia2018graphnetworks,gilmer2017mpnn}. The remainder provides a conceptual outline of the proposed framework, early validation results, and an outlook on future research directions.

\begin{figure*}[bt]
  \centering
  \includegraphics[width=0.95\linewidth]{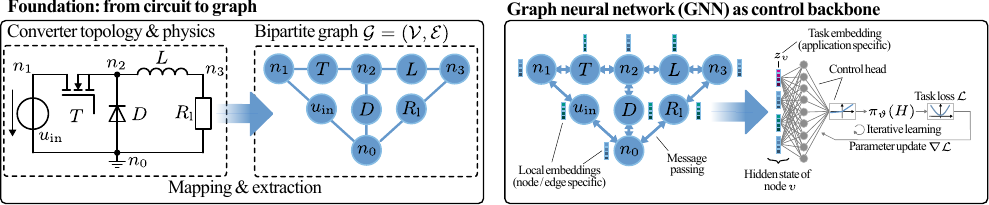}
  \caption{Abstracted high-level view of the proposed GNN-based  PEC control framework.}
  \label{fig:High_level_overview}
\end{figure*}

%==========================
\section{Graph Representations of PECs}
%==========================
\subsection{From netlists to typed bipartite graphs}
Let a PEC topology be represented by a typed bipartite graph
\begin{equation}
  \mathcal{G} \triangleq (\mathcal{V},\mathcal{E}),\qquad
  \mathcal{V}=\mathcal{V}_{\mathrm{C}}\cup\mathcal{V}_{\mathrm{N}},\qquad
  \mathcal{V}_{\mathrm{C}}\cap\mathcal{V}_{\mathrm{N}}=\emptyset,
  \label{eq:topology_graph}
\end{equation}
where \(\mathcal{V}_{\mathrm{C}}\) are component nodes (switches, passives, sources, loads, ports) and \(\mathcal{V}_{\mathrm{N}}\) are net/potential nodes. Each edge connects a component terminal to a net:
\begin{equation}
  \mathcal{E}\subseteq \mathcal{V}_{\mathrm{C}}\times \mathcal{V}_{\mathrm{N}}.
  \label{eq:edges}
\end{equation}
A subset \(\mathcal{V}_{\mathrm{S}}\subseteq\mathcal{V}_{\mathrm{C}}\) denotes controllable switching elements (i.e., transistors or entire legs/bridges). \(\mathcal{V}_{\mathrm{N}}\) corresponds to nets in a netlist (nodes of equal potential), and \(\mathcal{V}_{\mathrm{C}}\) corresponds to stamped components. Thus, \(\mathcal{G}\) is a graph-theoretic encoding of the converter schematic that naturally supports variable converter sizes/topologies.

To distinguish heterogeneous circuit primitives within a single graph model, we assign each node a discrete type label
\begin{equation}
  \mathrm{type}:\mathcal{V}\rightarrow \mathcal{T}_{\mathcal{V}},\qquad \abs{\mathcal{T}_{\mathcal{V}}}<\infty,
\end{equation}
where \(\mathcal{T}_{\mathcal{V}}\) is a finite set of node types (e.g., net node, transistor, diode, inductor, capacitor, source).
This map is (i) topology-intrinsic (derived from the netlist, not from node indexing) and (ii) used to select type-specific encoder and message/update functions in the GNN, so that physically distinct devices are processed by the appropriate parameter subset.

\subsection{Graph snapshots}
At discrete time \(t\in\N\), the controller input is a graph snapshot
\begin{equation}
  \mathcal{S}_t \triangleq \bigl(\mathcal{G},\,\bm{X}_t,\,\bm{E},\,\bm{z}_\tau\bigr),
  \label{eq:snapshot}
\end{equation}
with the following additional elements:
\begin{itemize}
  \item Node features: \(\bm{X}_t\in\R^{\abs{\mathcal{V}}\times d_x}\), with row vector \(\bm{x}_t(v)\in\R^{d_x}\) for each \(v\in\mathcal{V}\) and \(d_x\in\N_{>0}\). Features may include measured/estimated currents/voltages, operating-point indicators, and component parameters.
  \item Edge features: \(\bm{E}\in\R^{\abs{\mathcal{E}}\times d_e}\), with edge-feature vector \(\bm{e}(i,j)\in\R^{d_e}\) for each \((i,j)\in\mathcal{E}\), where \(d_e\in\N_{>0}\).
        Edge features can encode terminal index/port orientation, device polarity, or incidence metadata.
  \item Task/context features: \(\bm{z}_\tau\in\mathbb{Z}\subseteq\R^{d_z}\), \(d_z\in\N_{>0}\), where \(\mathbb{Z}\) is assumed compact.
        The task \(\tau\in\mathcal{T}\) may specify references (e.g., \(\bm{i}^*,\bm{v}^*\)), control objective weights (tracking vs.\ switching losses), constraint margins, and controller mode.
\end{itemize}

\subsection{Dynamical state \(\bm{y}\) and a measurement-to-graph embedding}
We denote the dynamical system state by \(\bm{y}_t\) (converter plus relevant load/grid states),
\[
  \bm{y}_t\in \mathbb{Y}_{\mathcal{G}}\subseteq \R^{n_y(\mathcal{G})},
  \qquad n_y(\mathcal{G})\in\N_{>0}.
\]
To map physical variables to graph features, define a (possibly engineered or learned) embedding map
\begin{equation}
  \bm{\mu}_{\mathcal{G}}:\mathbb{Y}_{\mathcal{G}}\times\mathbb{Z}\rightarrow \R^{\abs{\mathcal{V}}\times d_x},
  \qquad
  \bm{X}_t=\bm{\mu}_{\mathcal{G}}(\bm{y}_t,\bm{z}_\tau).
  \label{eq:mu_map}
\end{equation}
Here, \(\bm{\mu}_{\mathcal{G}}\) is a feature interface that assigns measured/estimated quantities to the appropriate nodes (e.g., inductor current to an inductor node, capacitor voltage to a capacitor node), enabling one encoder to operate across topologies.

\subsection{Permutation consistency and variable-size topologies}
A converter graph shall be storable with arbitrary node numbering. A meaningful encoder must therefore be independent of node labels. Let \(\Pi\) be any permutation of node indices that preserves the bipartite partition and types. A node-wise encoder \(\bm{\Phi}_{\bm{\theta}}\) is required to be permutation-equivariant:
\begin{equation}
  \bm{\Phi}_{\bm{\theta}}(\Pi\!\cdot\!\mathcal{G},\,\Pi\bm{X}_t,\,\Pi\bm{E},\,\bm{z}_\tau)
  =
  \Pi\,\bm{\Phi}_{\bm{\theta}}(\mathcal{G},\,\bm{X}_t,\,\bm{E},\,\bm{z}_\tau),
  \label{eq:equivariance}
\end{equation}
so that reindexing nodes only reindexes embeddings. Message-passing GNNs satisfy this property by construction when they use permutation-invariant aggregation \cite{battaglia2018graphnetworks,gilmer2017mpnn}.

\subsection{Task-conditioned message passing backbone}
Let \(L\in\N_{>0}\) be the number of message-passing layers. The backbone produces node embeddings
\(\bm{H}_t\in\R^{\abs{\mathcal{V}}\times d_h}\) and an optional global embedding \(\bm{h}_{\mathcal{G},t}\in\R^{d_g}\)
\begin{equation}
  \begin{gathered}
    (\bm{H}_t,\bm{h}_{\mathcal{G},t}) = \bm{\Phi}_{\bm{\theta}}(\mathcal{S}_t), \\
    \bm{\Phi}_{\bm{\theta}}:\mathfrak{S}\rightarrow
    \R^{\abs{\mathcal{V}}\times d_h}\times\R^{d_g},
    \quad
    \bm{\theta}\in\R^{n_\theta},
  \end{gathered}
  \label{eq:phi_def}
\end{equation}
where \(\mathfrak{S}\) denotes the set of admissible snapshots \eqref{eq:snapshot} and \(d_h,d_g\in\N_{>0}\).

We now define \(\bm{\Phi}_{\bm{\theta}}\) as typed bipartite message passing in two phases per layer (component\(\rightarrow\)net and net\(\rightarrow\)component). Initialize embeddings with a type-specific encoder:
\begin{equation}
  \bm{h}_t^{(0)}(v)
  =
  \bm{\mathrm{enc}}_{\mathrm{type}(v)}\!\left(\bm{x}_t(v),\bm{z}_\tau\right)\in\R^{d_h}
  \label{eq:enc}
\end{equation}
with $\bm{\mathrm{enc}}_{\alpha}:\R^{d_x}\times\R^{d_z}\to\R^{d_h}$ for each type $\alpha\in\mathcal{T}_{\mathcal{V}}$.

\paragraph{Phase A (component \(\rightarrow\) net)}
For each edge \((c,n)\in\mathcal{E}\) with \(c\in\mathcal{V}_{\mathrm{C}},n\in\mathcal{V}_{\mathrm{N}}\), define a message
\begin{equation}
  \bm{m}^{(\ell)}_t(c\!\to\!n)=
  \bm{\phi}^{(\ell)}_{\bm{\theta}}\!\left(
  \bm{h}^{(\ell)}_t(c),\bm{h}^{(\ell)}_t(n),\bm{e}(c,n),\bm{z}_\tau
  \right)\in\R^{d_m},
  \label{eq:msg_cn}
\end{equation}
where \(d_m\in\N_{>0}\) and
\[
  \bm{\phi}^{(\ell)}_{\bm{\theta}}:\R^{d_h}\times\R^{d_h}\times\R^{d_e}\times\R^{d_z}\to\R^{d_m}.
\]
Aggregate incoming messages at each net node \(n\) with a permutation-invariant aggregator
\begin{equation}
  \bm{a}^{(\ell)}_t(n)=
  \bm{\mathrm{AGG}}_{\mathrm{N}}\Bigl(\set{\bm{m}^{(\ell)}_t(c\!\to\!n):(c,n)\in\mathcal{E}}\Bigr)\in\R^{d_m},
  \label{eq:agg_n}
\end{equation}
where \(\bm{\mathrm{AGG}}_{\mathrm{N}}:\mathcal{M}(\R^{d_m})\to\R^{d_m}\) maps multisets of vectors to a vector (e.g., sum/mean/max).
Update the net embedding:
\begin{equation}
  \begin{gathered}
    \bm{h}^{(\ell+\frac12)}_t(n)=
    \bm{\psi}^{(\ell)}_{\bm{\theta}}\!\left(\bm{h}^{(\ell)}_t(n),\bm{a}^{(\ell)}_t(n),\bm{z}_\tau\right)\in\R^{d_h},\\
    \bm{\psi}^{(\ell)}_{\bm{\theta}}:\R^{d_h}\times\R^{d_m}\times\R^{d_z}\to\R^{d_h}.
  \end{gathered}
  \label{eq:upd_n}
\end{equation}
This resembles collecting contributions of incident components at a net, but learned and task-conditioned.

\paragraph{Phase B (net \(\rightarrow\) component)}
Similarly, messages from nets to a component \(c\in\mathcal{V}_{\mathrm{C}}\) are
\begin{equation}
  \bm{m}^{(\ell)}_t(n\!\to\!c)=
  \bm{\phi}^{(\ell)}_{\bm{\theta}}\!\left(
  \bm{h}^{(\ell+\frac12)}_t(n),\bm{h}^{(\ell)}_t(c),\bm{e}(c,n),\bm{z}_\tau
  \right)\in\R^{d_m},
  \label{eq:msg_nc}
\end{equation}
aggregated as
\begin{equation}
  \bm{a}^{(\ell)}_t(c)=
  \bm{\mathrm{AGG}}_{\mathrm{C}}\Bigl(\set{\bm{m}^{(\ell)}_t(n\!\to\!c):(c,n)\in\mathcal{E}}\Bigr)\in\R^{d_m},
  \label{eq:agg_c}
\end{equation}
and updated by
\begin{equation}
  \bm{h}^{(\ell+1)}_t(c)=
  \bm{\psi}^{(\ell)}_{\bm{\theta}}\!\left(\bm{h}^{(\ell)}_t(c),\bm{a}^{(\ell)}_t(c),\bm{z}_\tau\right)\in\R^{d_h}.
  \label{eq:upd_c}
\end{equation}
Repeat phases A and B for \(\ell=0,\dots,L-1\). After \(L\) layers of message passing, define \(\bm{H}_t\) by stacking \(\bm{h}^{(L)}_t(v)\). This enables a distributed control approach as the local GNN embedding \(\bm{h}^{(L)}_t(s)\) at a given switch node \(s\in\mathcal{V}_{\mathrm{S}}\) is a learned local descriptor of the switch’s electrical neighborhood and operating conditions. Optionally, a permutation-invariant readout yields a global embedding
\begin{equation}
  \begin{gathered}
    \bm{h}_{\mathcal{G},t}
    =
    \bm{r}_{\bm{\theta}}\Bigl(\set{\bm{h}^{(L)}_t(v):v\in\mathcal{V}}\Bigr)\in\R^{d_g},
    \\
    \bm{r}_{\bm{\theta}}:\mathcal{M}(\R^{d_h})\to\R^{d_g}.
  \end{gathered}
  \label{eq:global_readout}
\end{equation}
Here, \(\bm{h}_{\mathcal{G},t}\) summarizes global information and can be  utilized if a single central controller is targeted.

%==========================
\section{Optimal Control and Learning a Topology-Agnostic Meta-Controller}
%==========================
\subsection{General constrained finite-horizon optimal control problem}
For each PEC represented by a graph \(\mathcal{G}\), consider the discrete-time dynamics
\begin{equation}
  \bm{y}_{t+1}=\bm{f}_{\mathcal{G}}(\bm{y}_t,\bm{u}_t,\bm{d}_t),
  \label{eq:dynamics}
\end{equation}
where
\[
  \bm{y}_t\in\mathbb{Y}_{\mathcal{G}}\subseteq\R^{n_y(\mathcal{G})},\quad
  \bm{u}_t\in\mathbb{U}_{\mathcal{G}}\subseteq\R^{n_u(\mathcal{G})},\quad
  \bm{d}_t\in\mathbb{D}_{\mathcal{G}}\subseteq\R^{n_d(\mathcal{G})},
\]
are the system states, control inputs as well as external disturbances, while the system dynamics
\[
  \bm{f}_{\mathcal{G}}:\mathbb{Y}_{\mathcal{G}}\times\mathbb{U}_{\mathcal{G}}\times\mathbb{D}_{\mathcal{G}}\to\mathbb{Y}_{\mathcal{G}}
\]
are generally nonlinear (e.g., due to magnetic saturation or nonlinear loads). Constraints are expressed by
\begin{equation}
  \bm{g}_{\tau,\mathcal{G}}(\bm{y}_t,\bm{u}_t)\le \bm{0},
  \qquad
  \bm{g}_{\tau,\mathcal{G}}:\mathbb{Y}_{\mathcal{G}}\times\mathbb{U}_{\mathcal{G}}\to\R^{n_g(\tau,\mathcal{G})},
  \label{eq:constraints}
\end{equation}
elementwise, encoding protection limits (currents, voltages, duty bounds), and task-dependent envelopes.

Given horizon \(H\in\N_{>0}\), stage cost \(\ell_{\tau,\mathcal{G}}:\mathbb{Y}_{\mathcal{G}}\times\mathbb{U}_{\mathcal{G}}\to\Rge\) and terminal cost \(V_{\tau,\mathcal{G}}:\mathbb{Y}_{\mathcal{G}}\to\Rge\), define the finite-horizon optimal control problem (OCP):
\begin{align}
  \min_{\{\bm{u}_t,\dots,\bm{u}_{t+H-1}\}}
  \quad            &
  \sum_{k=0}^{H-1}\ell_{\tau,\mathcal{G}}(\bm{y}_{t+k},\bm{u}_{t+k})
  +
  V_{\tau,\mathcal{G}}(\bm{y}_{t+H})
  \label{eq:ocp}                                                                        \\
  \text{s.t.}\quad &
  \bm{y}_{t+k+1}=\bm{f}_{\mathcal{G}}(\bm{y}_{t+k},\bm{u}_{t+k},\bm{d}_{t+k}),\nonumber \\
                   &
  \bm{g}_{\tau,\mathcal{G}}(\bm{y}_{t+k},\bm{u}_{t+k})\le\bm{0},\nonumber               \\
                   &
  \bm{u}_{t+k}\in\mathbb{U}_{\mathcal{G}},\quad \bm{y}_{t}\ \text{given.}\nonumber
\end{align}
While the OCP structure is defined traditionally, the challenge is that \(\mathcal{G}\) (PEC topology and parameters) as well as \(\tau\) (objective/constraints) may vary across applications.

\subsection{Topology-agnostic meta-control policy}
To overcome this challenge, we define a meta-policy that maps a snapshot to an admissible control action:
\begin{equation}
  \bm{u}_t = \bm{\pi}_{\bm{\vartheta}}(\mathcal{S}_t),
  \qquad
  \bm{\pi}_{\bm{\vartheta}}:\mathfrak{S}\to \bigcup_{\mathcal{G}}\mathbb{U}_{\mathcal{G}},
  \qquad
  \bm{\vartheta}\in\R^{n_\vartheta}.
  \label{eq:policy}
\end{equation}
Using the backbone \eqref{eq:phi_def}, we implement a backbone plus control head decomposition
\begin{equation}
  (\bm{H}_t,\bm{h}_{\mathcal{G},t})=\bm{\Phi}_{\bm{\theta}}(\mathcal{S}_t),
  \qquad
  \bm{u}_t=\bm{\Gamma}_{\bm{\omega}}(\bm{H}_t,\bm{h}_{\mathcal{G},t},\bm{z}_\tau),
  \label{eq:backbone_head}
\end{equation}
with $\bm{\vartheta}=(\bm{\theta},\bm{\omega})$. Here, \(\bm{\Gamma}_{\bm{\omega}}\) is the control head, mapping embeddings to converter actuation. Above, the global context $\bm{h}_{\mathcal{G},t}$ is optional and may be omitted for fully distributed control: Let each controllable element \(s\in\mathcal{V}_{\mathrm{S}}\) have a local actuation vector \(\bm{u}_t(s)\in\mathbb{U}_s\subseteq\R^{d_u(s)}\), where \(d_u(s)\in\N_{>0}\) may encode a single or a subset of transistors, e.g., within a PEC leg. Define
\begin{equation}
  \begin{gathered}
    \bm{u}_t(s)=
    \bm{\gamma}_{\bm{\omega}}\!\left(\bm{h}^{(L)}_t(s),\bm{z}_\tau\right)\in\mathbb{U}_s,\\
    \bm{\gamma}_{\bm{\omega}}:\R^{d_h}\times\R^{d_z}\to\R^{d_u(s)}.
  \end{gathered}
  \label{eq:switch_head}
\end{equation}
The full action vector for the entire PEC is then the concatenation \(\bm{u}_t=\mathrm{concat}\bigl(\bm{u}_t(s)\bigr)_{s\in\mathcal{V}_{\mathrm{S}}}\in\mathbb{U}_{\mathcal{G}}\). In this distributed fashion, the same head \(\bm{\gamma}_{\bm{\omega}}\) is applied to each switch node, so the policy naturally scales to converters with different numbers of switches while sharing the same backbone and head parameters.

\subsection{End-to-end DPC training}
Differentiable predictive control (DPC) trains \(\bm{\pi}_{\bm{\vartheta}}\) by minimizing a multi-step control objective while differentiating through a differentiable plant/surrogate model \(\bm{f}_{\mathcal{G}}\)  using automatic differentiation \cite{drgona2022dpc,li2026dpc_tpel}. The key benefit is numerical amortization: instead of solving \eqref{eq:ocp} online at each \(t\), we learn a fast policy that approximates the optimal decision rule over a distribution of scenarios.

Let \((\mathcal{G},\tau)\sim p(\mathcal{G},\tau)\), initial states \(\bm{y}_t\sim p_0(\cdot\mid\mathcal{G},\tau)\), and disturbances \(\bm{d}_{t:t+H-1}\sim p_d(\cdot\mid\mathcal{G},\tau)\).
Define the closed-loop rollout for \(k=0,\dots,H-1\):
\begin{equation}
  \begin{gathered}
    \bm{X}_{t+k}=\bm{\mu}_{\mathcal{G}}(\bm{y}_{t+k},\bm{z}_\tau),\quad
    \mathcal{S}_{t+k}=(\mathcal{G},\bm{X}_{t+k},\bm{E},\bm{z}_\tau),\\
    \bm{u}_{t+k}=\bm{\pi}_{\bm{\vartheta}}(\mathcal{S}_{t+k}),\quad
    \bm{y}_{t+k+1}=\bm{f}_{\mathcal{G}}(\bm{y}_{t+k},\bm{u}_{t+k},\bm{d}_{t+k}).
  \end{gathered}
  \label{eq:rollout}
\end{equation}
A soft-constrained training objective is
\begin{align*}
  \begin{split}
    \min_{\bm{\vartheta}\in\R^{n_\vartheta}}\quad
    \mathcal{L}(\bm{\vartheta})
    \triangleq
    \E\Bigg[
      \sum_{k=0}^{H-1}\ell_{\tau,\mathcal{G}}(\bm{y}_{t+k},\bm{u}_{t+k})
      +V_{\tau,\mathcal{G}}(\bm{y}_{t+H})
    \\+\lambda\sum_{k=0}^{H-1}\varphi\!\left(\bm{g}_{\tau,\mathcal{G}}(\bm{y}_{t+k},\bm{u}_{t+k})\right)
      \Bigg],
  \end{split}
  \label{eq:dpc_loss}
\end{align*}
where \(\lambda\in\R_{\ge 0}\) and \(\varphi:\R^{n_g}\to\R_{\ge 0}\) is differentiable (e.g., squared hinge using \(\relu\)). Deployment-time safety layers (e.g., predictive safety filters) can be added to enforce hard constraints if needed \cite{wabersich2021psf}. Solving \eqref{eq:ocp} via DPC (in an end-to-end differentiable computing environment, e.g., Python/JAX) yields a topology-agnostic meta-controller that generalizes across converter graphs and tasks. However, alternative solutions using the same graph-conditioned policy class, like reinforcement learning  \cite{wang2018nervenet} or imitation learning \cite{novak2021imitation}, are also possible.

\section{Early validation}
Following the buck converter application from Fig.~\ref{fig:High_level_overview}, we randomly sample 100 different buck converter configurations by varying \(L\in\SIrange{1e-7}{2e-1}{\henry}\), \(C\in\SIrange{5e-8}{2e-2}{\farad}\), and \(R_\mathrm{l}\in\SIrange{1e-2}{1e3}{\ohm}\). A GNN-based meta-controller is trained via DPC over a distribution of reference-tracking tasks (output voltage reference step changes) using a differentiable PEC simulation openly released in \cite{gnn_buck_converter_control}. As cost function in \eqref{eq:ocp}, we use a mean squared error (MSE) voltage tracking error.

\begin{figure}[t]
  \centering
  \subfloat[Example case 1\label{fig:tpel080_buck25}]{%
    \includegraphics[width=0.95\linewidth,height=0.32\textheight,keepaspectratio]{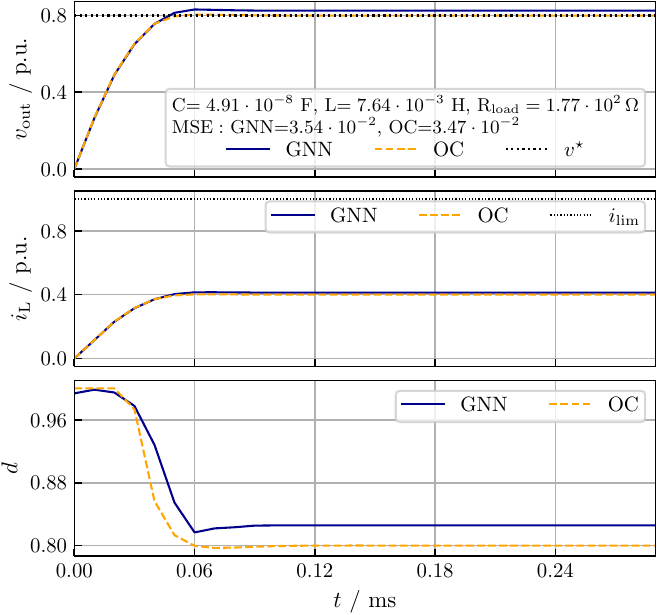}%
  }\par\smallskip
  \subfloat[Example case 2\label{fig:tpel080_buck75}]{%
    \includegraphics[width=0.95\linewidth,height=0.32\textheight,keepaspectratio]{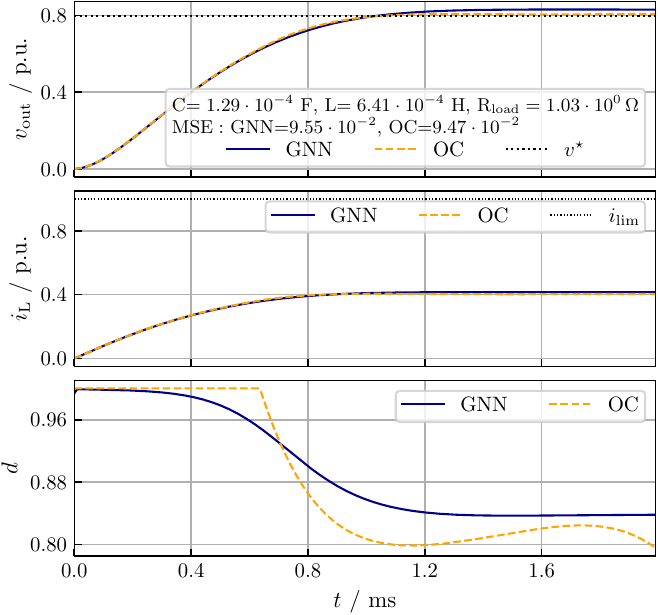}%
  }
  \caption{Time series of two exemplary buck converters controlled by the same GNN-based meta-controller and comparison to the achievable optimal control (OC) performance.}
  \label{fig:validation_tpel080_buck}
\end{figure}

Fig.~\ref{fig:validation_tpel080_buck} compares the trained GNN-based meta-controller with an optimal-control (OC) baseline that solves \eqref{eq:ocp} online via nonlinear programming with full model knowledge for each validation case independently, thus approximating the achievable performance bound for a given configuration. In both exemplary test cases, the learned controller attains near-optimal tracking despite being trained across a broad range of converter parameters and operating points.
\begin{figure}[t]
  \centering
  \subfloat[Box plot of closed-loop performance \label{fig:box_plot}]{%
    \includegraphics[width=0.7\linewidth]{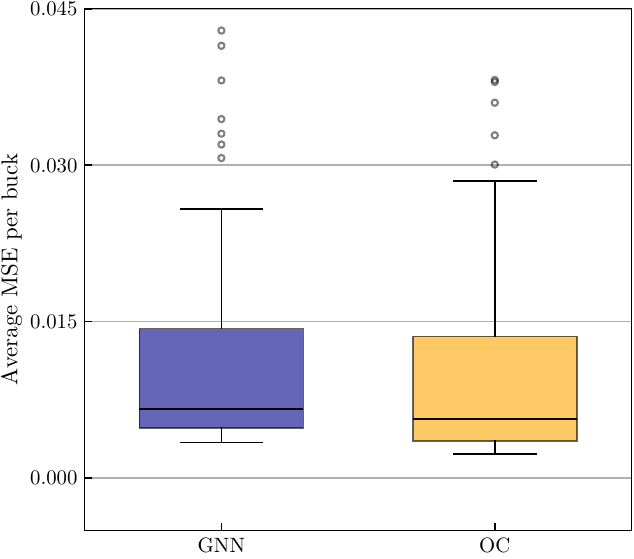}%
  }\par\smallskip
  \subfloat[Histogram of relative performance differences\label{fig:relative_hist}]{%
    \includegraphics[width=0.85\linewidth]{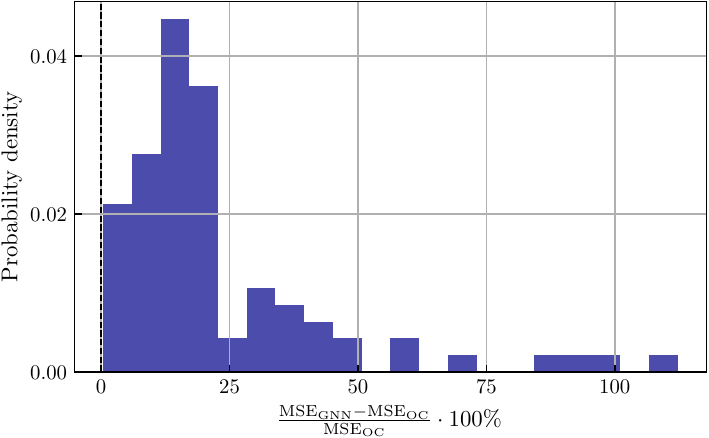}%
  }
  \caption{Closed-loop control performance comparison of 100 randomly sampled buck converters considering 10 step-response reference tracking cases each. }
  \label{fig:sampled_performance}
\end{figure}

Fig.~\ref{fig:sampled_performance} evaluates a representative set of buck-converter configurations and operating-point variations. The GNN-based meta-controller delivers close-to-optimal performance across all samples, with a median relative gap of 16.7\% to the achievable optimum. Since we have not yet performed hyperparameter optimization of the GNN architecture/training procedure nor applied domain-specific feature engineering, these early results remain highly promising and motivate further research.

%==========================
\section{Conclusion and outlook}
%==========================
We proposed a meta-controller that uses a GNN backbone for PEC control and demonstrated proof-of-concept feasibility via near-optimal performance on diverse buck converter configurations. Although the validation covered only one topology, the results support the approach's feasibility as a first step toward a general framework. Future work will extend the method to heterogeneous converter families, broader objectives and constraint sets, and will study sim-to-real transfer and real-time embedded deployment, including scalability, interpretability, and robustness to model mismatch and unmodeled dynamics.

\bibliographystyle{IEEEtran}
\bibliography{refs}

\end{document}